# Time-of-flight discrimination between gamma-rays and neutrons by neural networks

Serkan AKKOYUN[a,*]

[a]*Faculty of Science, Department of Physics, Cumhuriyet University, Sivas, Turkey*

**Abstract**

In gamma-ray spectroscopy, a number of neutrons are emitted from the nuclei together with the gamma-rays and these neutrons influence gamma-ray spectra. An obvious method of separating between neutrons and gamma-rays is based on the time-of-flight (*tof*) technique. This work aims obtaining *tof* distributions of gamma-rays and neutrons by using feed-forward artificial neural network (ANN). It was shown that, ANN can correctly classify gamma-ray and neutron events. Testing of trained networks on experimental data clearly shows up *tof* discrimination of gamma-rays and neutrons.

*Keywords:* Artificial neural network, time-of-flight, HPGe, Monte Carlo simulation.

[*]Corresponding author: E-mail: sakkoyun@cumhuriyet.edu.tr, Tel: +903462191010x1413

## 1. Introduction

In nuclear structure physics studies, heavy-ion fusion-evaporation (HIFE) reactions, which are widely used type of nuclear reactions, are invaluable tool. In these types of reactions, two nuclei fuse to form a compound nucleus. This nucleus decays by evaporating numerous light particles, principally neutrons, protons and alpha particles. Contrary to the charged ones, neutrons can travel long distances and interact in the detectors together with the gamma-rays. These neutrons cause unwanted background in gamma-ray spectra. In the germanium detectors, used in this study, the main energy deposition mechanisms of neutrons following HIFE reactions are elastic and inelastic scatterings. Inelastic scatterings of neutrons are more complex due to the fact that the excitation of the recoiling germanium nuclei.

Time-of-flight (*tof*) method is an obvious way to discriminate gamma-ray and neutron events in the detectors. In this work, in order to obtain *tof* distributions of the gamma-rays and the neutrons, which are considered to be emitted from a HIFE reaction, Geant4 (Agostinelli et al, 2003) Monte Carlo simulations are performed. After that, these distributions are also obtained by using artificial neural network (ANN) method. Since recently, ANNs have been successfully used in many fields including discrimination of neutrons and gamma-rays (Cao et al, 1998; Esposito et al, 2004; Liu et al, 2009; Akkoyun and Yildiz, 2012). It will be clearly seen that the ANN method, which has speed advantage, is consistent with the experimental results. By training on the simulation data belonging only to the gamma-rays or neutrons, test set predictions were performed to observe *tof* discrimination between neutrons and gamma-rays.

## 2. Monte Carlo Simulations of HPGe detectors

The simulations of HPGe detector and the interactions of gamma-rays and neutrons inside this detector were simulated under Geant4.9.2 Monte Carlo simulation program (Agostinelli et al, 2003). The diameter and thickness of the cylindrical planar detector were 7 and 1.5 cm respectively. The detector consists of high purity germanium (HPGe) crystals. The distance between gamma-ray/neutron source and the front surface of the detector was 25 cm.

In the simulations, gamma-rays, neutrons and both of them together were sent to detector from the source position. The energies of the incident neutrons with 4 multiplicities were sampled from a distribution (at 0-15 MeV interval) which is obtained from a typical HIFE reaction. The flight times are 58 ns for 100 keV neutrons and 4.5 ns for 15 MeV neutrons to the front surface of the detectors. The discrete energies of the gamma-ray cascade with multiplicity 30 are in 1-3 MeV interval. The flight times of the gamma-rays are around 1 ns. In Fig.1, *tof* distributions of gamma-rays and neutrons were given together. This study aims to give an alternative way to obtain *tof* distributions of gamma-rays and neutrons. Using these informations can help classification and discrimination of the gamma-ray and neutron events by using artificial neural network method.

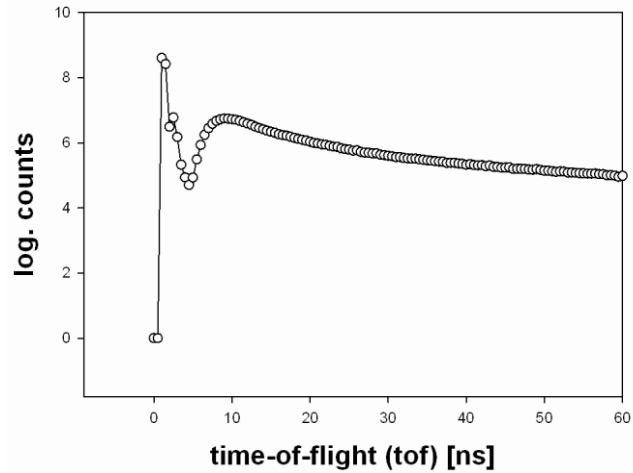

**Fig.1.** Time-of-flight (*tof*) distributions of gamma-rays and neutrons in HPGe detectors.

## 3. Artificial Neural Networks

The main task of the artificial neural networks (ANNs) is to give outputs in consequence of the computation of the inputs. ANNs are mathematical models that mimic the human brain. They consist of several processing units called neurons which have adaptive synaptic weights (Haykin, 1999). ANNs are also effective tools for pattern recognition. The classical ANN consists of three layers: input, hidden and output (Fig.2). The number of hidden layers can differ, but a single hidden layer is enough for efficient nonlinear function approximation (Hornik et al, 1989). In this study, one input layer with one neuron, one hidden layer with many (*h*) neuron and one output layer with one neuron (*1-h-1*) ANN topology was used for accurately and reliably prediction of *tof* distributions of the gamma-rays and neutrons. Analyses were performed for different hidden neuron numbers, $h=$ 6, 15 and 25. The total numbers of adjustable weights were 12, 30 and 50 according to the formula given in Eq. (1):

$$(p \times h + h \times r = h \times (p+r) = 2h) \tag{1}$$

where $p$ and $r$ are the input and output neuron numbers, respectively.

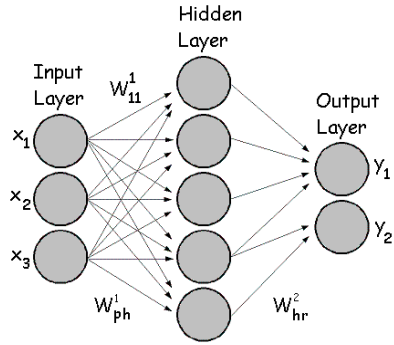

**Fig.2.** Fully connected one input-many hidden-one output layer LFNN. Only one hidden layers are shown. $x_i$ and $y_i$ are input and output vector components respectively. Circles are neurons and arrows indicate adaptable synaptic weights. $w^i_{jk}$: weight vector component, where i is a layer index, jk weight component from the *j*th neuron of *i*th layer and to *k*th neuron of *(i+1)*th layer.

The neuron in the input layer collects the data from environment and transmits via weighted connections to the neurons of hidden layer which is needed to approximate any nonlinear function. The hidden neuron activation function can be theoretically any well-behaved nonlinear function. The type of activation function was chosen as hyperbolic tangent for hidden layer (Eq. (2)):

$$tanh = \frac{(e^x - e^{-x})}{(e^x + e^{-x})} \tag{2}$$

Note that instead of Eq. (2), any other suitable sigmoidal function could also be used. The output layer neurons return the signal after the analysis.

The ANN train and test datasets used in this work were produced by Geant4 simulations which are mentioned in Section 2. At the first step of the simulations, gamma-ray experimental data (dataset-I) and neutron experimental data (dataset-II) were generated separately mainly for training stage. In the last simulation, the data (dataset-III) including both gamma-rays and neutrons together were generated for application of the ANN method.

An ANN software NeuroSolutions v6.02 was also used. The ANN input was *tof* values of the gamma rays and/or the neutrons and the desired output was detector response for *tof*. In the training step of this study, dataset-I and dataset-II were used separately. For all ANN processing case, dataset-I and II were divided into two separate sets. One of this is for the training step (about 50% of all data) and the rest is for the test step. In the training step, a back-propagation algorithm with Levenberg-Marquardt for the training of the ANN was used. The maximum epoch number (one complete presentation of the all input-output data to the network being trained) was 1000. By convenient modifications, ANN modifies its weights until an acceptable error level between predicted and desired outputs is attained. The error function which measures the difference between outputs was *mean square error (MSE)* as given in Eq. (3):

$$MSE = \frac{[\sum_{k=1}^{r}\sum_{i=1}^{N} y_{ki} - f_{ki})^2]}{N} \qquad (3)$$

where N is the number of training or test samples, whichever applies, $y_{ki}$ and $f_{ki}$ are the desired output and network output, respectively. Then by using ANN with final weights, the performance of the network is tested over an unseen data. If the predictions of the test dataset are good enough, the ANN is considered to have consistently learned the functional relationship between input and output data (Yildiz, 2005). In this study for different *h* numbers, the mean of minimum MSE values for training were about $8 \times 10^{-27}$ for the gamma-rays and $2 \times 10^{-5}$ for the neutrons. In Fig.3, the training MSE values versus epoch number were given.

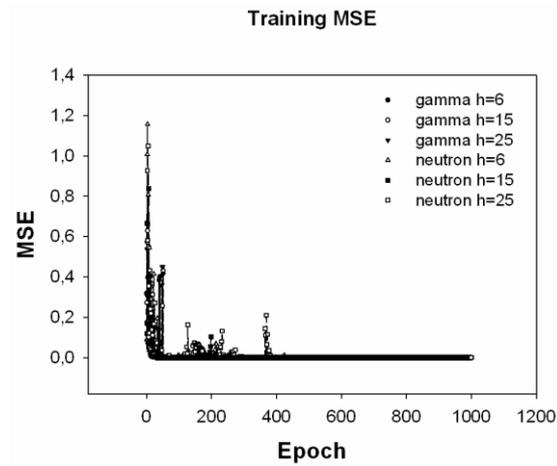

**Fig.3.** For gamma-ray and neutron datasets, the training MSE values versus epoch number for different *h* values.

## 4. Results and discussions

*4.1 Training and testing*

After the training of the networks on the half of the dataset-I and II separately with different hidden neuron numbers ($h = 6$, 15 and 25), trained networks were tested over the rest of the datasets which have never seen before by the networks. For the *tof* distributions of the gamma rays and the neutrons, the ANN predictions corresponding to dataset-I and II were given in Fig. 4 and 5, respectively. As can be clearly seen in the figures, the ANN predictions (*nno*) agree exceptionally well with highly nonlinear experimental (exp) values for each different *h* number. It is also apparently noticeable that, there are no significant changes in predictions with varying *h* numbers. Results obviously indicate that the test set ANN of the detector responses versus *tof* distributions has consistently generalized the train set ANN fittings.

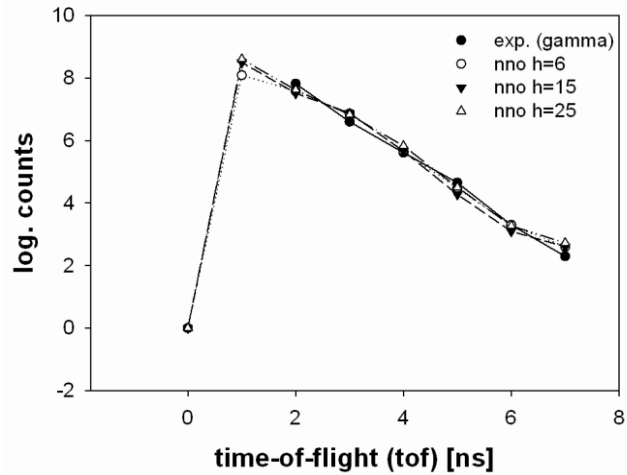

**Fig.4.** Experimental (*exp*) and ANN output (*nno*) test set predictions for *tof* distributions of the gamma-rays for different hidden layer neuron numbers ($h$).

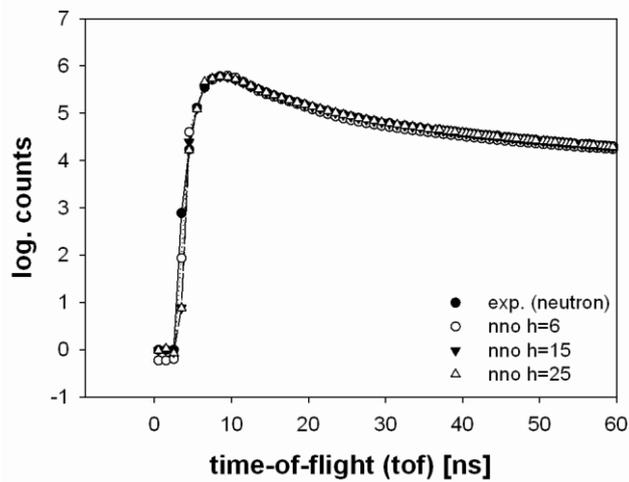

**Fig.5.** Experimental (*exp*) and ANN output (*nno*) test set predictions for *tof* distributions of the neutrons for different hidden layer neuron numbers ($h$).

*4.2 Classification and discrimination of the events*

In the event classification stage of the work, we let the output of the ANN be 1 or -1 according to whether input is gamma-ray or neutron events, respectively. Namely, if the input belongs to the dataset-I, the output was set to 1 and if the input belongs to the dataset-II, the output was set to -1. After training of the network under these conditions, ANN was tested over dataset-III including gamma-ray and neutron events together for classification. As indicated in Fig. 6, the networks classify different events with great success. So, the gamma-ray events are located correctly in the right-hand side of the y-axis and the neutron events are located in the left-hand side of the y-axis. Besides this, some events indicated by triangle symbol in the figure, are related to either gamma-ray or neutron events. Therefore, any event can be accepted with bigger probability as gamma-ray or neutron event according to the location in the side of the y-axis.

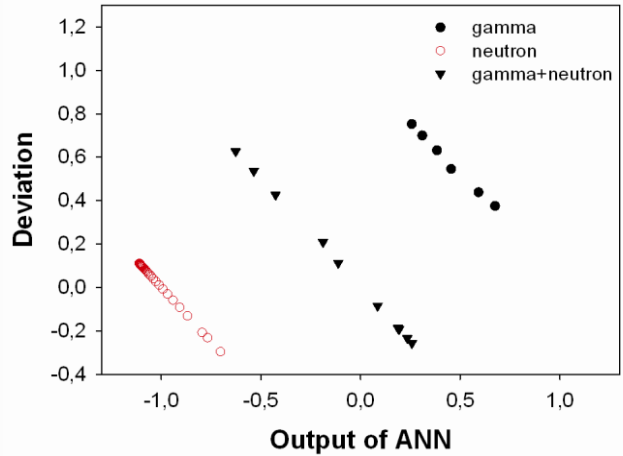

**Fig.6.** Classification of the gamma-ray and neutron events by using ANN. The filled circles represent gamma-rays and open circles represent neutrons. The triangle symbols are used for the events belonging either gamma-ray or neutrons.

In the last step of this work, the ANN method had been applied to discriminate *tof* distributions of the gamma-ray and neutron events. Firstly, the network which is trained by dataset-I was applied over experimental data (dataset-III) including gamma-ray and neutron events together. As shown in Fig. 7, the network ignores the neutron events by doing predictions corresponding to the gamma-ray events. In the figures, abbreviation *nno* refers to ANN outputs for predictions. Contrarily, if dataset-II was used for training, the network will ignore, this time, the gamma-ray events by doing predictions corresponding to the neutron events (Fig. 8). All the analysis in this stage were performed for constant hidden neuron number ($h=6$) because of varying $h$ number does not significantly alter the results. These predictions can help discrimination between gamma-ray and neutron events. As can be determined in the figures, the cut value on the distributions is about 4.5 ns for separation of the gamma-rays and the neutrons.

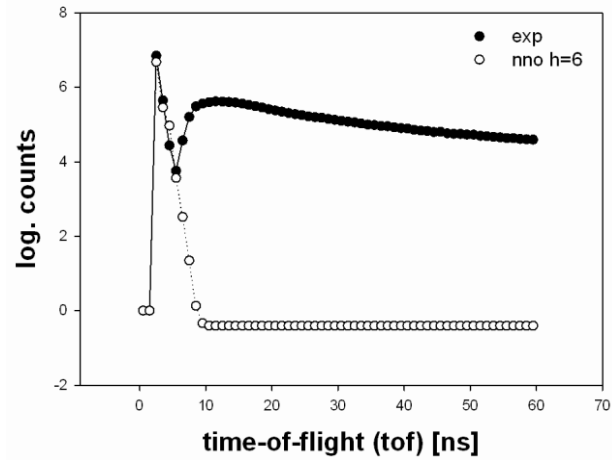

**Fig.7.** Experimental (*exp*) and ANN output (*nno*) test set predictions with *h*=6 for *tof* discrimination between gamma-rays and neutrons. The training of the ANN was done by using dataset-I.

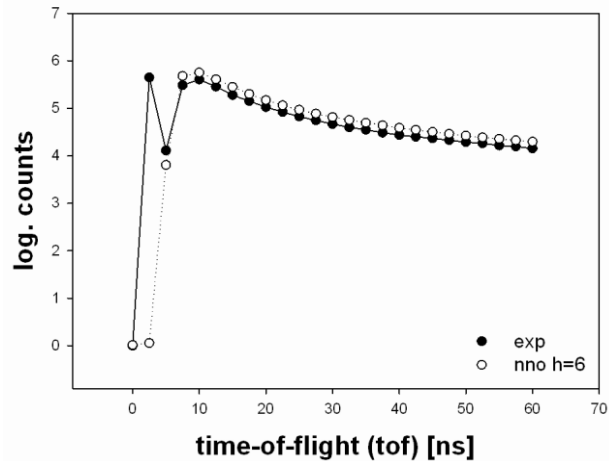

**Fig.8.** Experimental (*exp*) and ANN output (*nno*) test set predictions with *h*=6 for *tof* discrimination between gamma-rays and neutrons. The training of the ANN was done by using dataset-II.

## 5. Conclusions and potential applications

In this paper, we generated *tof* distributions in HPGe detectors for gamma-rays and neutrons both experimentally and by using artificial neural networks. These distributions help separation between gamma-ray and neutron events. It was clearly seen that the ANN method, which can be apply very fast, was consistent with the experimental results. The advantage of the ANN method is that it does not need any relationship between input and output data. It was also seen that ANN method is very successful in event classification. In this study, by training on the simulation data belonging only to the gamma-rays or neutrons, test set predictions were performed to observe *tof* discrimination between gamma-rays and neutrons. According to the simulations, the cut value on the distribution for *tof* separation was determined as about 4.5 ns. This value is useful for eliminating neutron background in the gamma-ray spectra generated in this work. But such an analysis is out of the scope of this work.

**Acknowledgments**

This work was supported by Cumhuriyet University (BAP Proj. no. F-361).